\documentclass[floatfix,aps,prl,showpacs,amsmath,nofootinbib,
preprintnumbers,twocolumn]{revtex4}
\usepackage{graphicx}
\usepackage{dcolumn}
\usepackage{bm}
\def\be{\begin{equation}}
\def\ee{\end{equation}}
\def\bea{\begin{eqnarray}}
\def\eea{\end{eqnarray}}
\newcommand{\Hu}{{\cal H}} 
 
\newcommand{\dd}{{\rm d}}

\newcommand{\Pk}{{\Phi_k}\!\!}

\newcommand{\nS}{n_{_{\rm S}}}

\newcommand{\AS}{A_{_{\rm S}}}
\newcommand{\lP}{\ell_{_{\rm Pl}}}

\begin{document}


\title{Spectra of primordial fluctuations in two-perfect-fluid
regular bounces}

\author{Fabio Finelli}
\email{finelli@iasfbo.inaf.it}
\affiliation{INAF/IASF-Bologna,
Istituto di Astrofisica Spaziale e
Fisica Cosmica di Bologna, Istituto Nazionale di Astrofisica,
via Gobetti, 101 -- I-40129 Bologna -- Italy}
\affiliation{
INAF/OAB, Osservatorio Astronomico di Bologna,
Istituto Nazionale di Astrofisica,
via~Ranzani~1, 40127~Bologna, Italy} 
\affiliation{
INFN, Istituto Nazionale di Fisica Nucleare, Sezione di Bologna, via 
Irnerio~46, 40126~Bologna, Italy}

\author{Patrick Peter}
\email{peter@iap.fr}
\affiliation{Institut d'Astrophysique de Paris - GreCo, FRE 2435-CNRS,
98bis boulevard Arago, 75014 Paris, France}

\author{Nelson Pinto-Neto}
\email{nelsonpn@cbpf.br}
\affiliation{Centro Brasileiro de Pesquisas Fisicas, Rue Dr. Xavier
Sigaud 150, Urca 22290-180, Rio de Janeiro, RJ, Brazil}

\date{\today}

\begin{abstract}
We introduce analytic solutions for a
class of two components bouncing models, where the bounce is
triggered by a negative energy density perfect fluid. The
equation of state of the two components are constant in time,
but otherwise unrelated. By numerically integrating
regular equations for scalar cosmological perturbations,
we find that the (would be) growing mode of
the Newtonian potential before the bounce never
matches with the the growing mode in the expanding stage.
For the particular case of a negative energy density component with a stiff
equation of state we give a
detailed analytic study, which is in complete agreement with the numerical
results. We also perform analytic and numerical calculations for 
long wavelength tensor perturbations, obtaining that, in most cases of interest,
 the tensor 
spectral index is independent
of the negative energy fluid and given by the spectral index of the 
growing mode in the contracting stage. 
We compare our results with previous investigations in the literature.
\end{abstract}


\maketitle

\section{Introduction}

Cosmological models with a bounce \cite{bounce} - a contraction which
reverses into an expansion - may solve the horizon problem \cite{GV}
in a non-inflationary way, i.e. as the Pre Big-Bang \cite{review} and
Ekpyrotic \cite{ekp} scenarios. In order to make bouncing models real
competitors of inflation, or at least complementary to it, while
addressing the singularity problem unavoidably present in such models,
the spectrum of density perturbations which results from the bounce
should be understood as it is in inflationary theories. Unfortunately,
the physics of cosmological perturbations during a bounce is much more
subtle, because of the reversal of growing and decaying modes before
and after the bounce, so that even though the bounce duration itself
may be very short, usual matching conditions \cite{DM,BF} should be used
with particular care and verified.

These subtleties have generated many works on the subject, in
particular after the proposal of the Ekpyrotic scenario \cite{ekp},
which is based on a very slow contraction and needs a bounce, as in the
Pre Big-Bang model studied
in the Einstein frame. Unanimous conclusions on
the resulting spectrum of metric fluctuations in the expanding stage
are still to come. Among the ongoing controversies, one concerned
the fate of cosmological
perturbations in a hydrodynamical radiation bounce triggered by a
negative energy density scalar field \cite{PPN}, generalized
afterwards in \cite{F}. This bounce - and its generalization - has
been suggested as a simple toy model, which has the advantage of
providing analytic solutions for the background, although having a
component which violates the null energy condition. Note that without
assuming spatial curvature and demanding general relativity to hold,
such a negative energy component is required at the level of an
effective theory in order for a bounce to take place \cite{PPN27}.

The initial result in this class of two-fluid models was that the spectrum of
the Newtonian potential after the bounce was the same as that of the
growing mode before the bounce. Such a result was obtained evolving
numerically \cite{PPN} and analytically \cite{PPN,F} a set of regular
equations. This result was later challenged in \cite{BV1,BV2}, generalized in
Ref.~\cite{VB} in which the scalar perturbations are evolved through a bounce 
characterized by a single physical scale, arguing
that the growing mode before the bounce matched only with the decaying
mode after the bounce, a possibility which has been already found
\cite{DM,BF,FB,TBF,CC}. However, the analysis of \cite{BV1,BV2}, demanding
the most general possible situation (the case at hand in the present work
being a subset), needs to rely on a set of singular equations, a fact that
could cast doubts on its accuracy had they use them directly; these
authors however obtained the solution in the form of a Born-like series
containing only convergent integrals; Non-singular equations have also been
evolved in different contexts, e.g. with a double scalar field bounce \cite{AW}
or one with a non-local dilaton potential stemming from string theory \cite{GGV}.
Note that for tensor modes, we already know that a matching between growing
modes before and after the bounce occurs~\cite{F,PPPN}.

In this paper we consequently reanalyze the behaviour of cosmological
perturbations during the radiation bounce, obtaining results in
agreement with these later studies \cite{BV1,BV2}, and in
contrast with our previous findings for scalar perturbations 
in \cite{PPN,F}. In section II we
present the background bouncing models containing
two perfect fluids with linear and unrelated equations of state. We 
also show how to describe 
the negative energy perfect fluid in terms of a K-essence scalar field.
In section III we propose a set of 
regular equations for linear perturbations of the above background models, 
one for the
Newtonian potential and the other for the velocity potential of the
fluid responsible for the bounce, which can be related to the linear
perturbations of the K-essence
scalar field yielding simpler regular equations suitable for the numerical
analysis presented in section IV  (another set involving the density contrast 
instead of the velocity potential is given in the appendix). 
In section V we justify some of the numerical results
through an analitical study of approximate solutions and their
matchings. We end in section VI with discussions and 
conclusions.

\section{Background}

We shall consider a class of
bouncing universes filled by two non-interacting perfect fluids with parameters
of state $w_+ \,, w_-$, {\em constant} in time \cite{F}, relating the energy densities $\epsilon_\pm$
to the pressures $p_\pm$ through $p_{\pm}=w_{\pm}\epsilon_{\pm}$.
The Einstein energy constraint for homogeneous and isotropic 
solutions is
\be
H^2 = \left( \frac{\dd a}{a \dd t} \right)^2
=\lP^2 \left[ \frac{\rho_+}{a^{3(1+w_+)}} -
\frac{\rho_-}{a^{3(1+w_-)}} \right] \,,
\label{hubble}
\ee
where $8 \pi G = M_{\rm pl}^{-2} = 3 \lP^2$, $\rho_+$, $\rho_-$ being
constants. Eq. (\ref{hubble}) is obtained using the background FLRW metric
\be
\dd s^2 = a^2(\eta) \left( \dd\eta^2 - \delta_{ij} \dd x^i \dd x^j \right),
\label{dg0}
\ee
and by  assuming energy conservation for each
single fluid separately in order to make explicit its dependence on the scale factor.
We restrict ourselves to $w_- > w_+$. It is clear that the negative energy density fluid,
or, in other words, $\rho_-$, is important only close to the bounce, in agreement with
what is required from a phenomenological model.

By introducing a new coordinate time $\tau$:
\be
\dd \tau = \frac{\dd t}{a^\beta} \,, \quad
{\rm with} \quad \beta = \frac{3}{2} (2 w_+ - w_-+1),
\label{newtime}
\ee
we can solve Eq.~(\ref{hubble}) for the scale factor as
\be
a (\tau) = a_0 \left( 1 + \frac{\tau^2}{\tau_0^2} \right)^\alpha \,,
\label{sfevolution}
\ee
with
\begin{eqnarray}
\alpha &=&  \frac{1}{3 (w_- - w_+)}, \\
a_0 &=& \left( \frac{\rho_-}{\rho_+} \right)^\alpha ,\\
\tau_0^2 &=& \frac{4\alpha^2}{\lP^2} \frac{\rho_-}{\rho^2_+}.
\end{eqnarray}
Note that the new coordinate time $\tau$ makes it possible
to generalize the solution, obtained in terms of the usual conformal
time found in \cite{F} to arbitrary values of $w_+ \,, w_-$.
Note also that this new coordinate time allowed to get general solutions
for the scale factor in a universe filled by
dust plus dark energy, the latter having an arbitrary constant equation of state \cite{GF}.

One can also describe fluids in terms of velocity potentials
\cite{schutz}. In the case of a perfect fluid, the velocity potential
action is very simple, and identical to an action for a
K-essence scalar field \cite{kessence} (which, incidentally, can also be used
to produce a bounce with no curvature and only one scalar degree of
freedom \cite{kbounce}), namely
\begin{equation}
{\cal S} =\int \left[\pm {\frac12} (\nabla_\mu \phi \nabla^\mu
\phi)^{(1+w_{\pm})/(2w_{\pm})} \right] \sqrt{-g}\, \dd^4x ,\label{actionK}
\end{equation}
where the $\pm$ sign is chosen according to whether the fluid has positive or
negative energy density. The energy-momentum tensor for $\phi$ reads
\begin{eqnarray}
T_{\mu\nu} &=& \pm \biggl[\frac{(1+w_{\pm})}{2w_{\pm}} \nabla_\mu \phi \nabla_\nu\phi
(\nabla_\beta \phi \nabla^\beta\phi)^{(1-w_{\pm})/(2w_{\pm})} \nonumber \\
&-&\frac{1}{2}g_{\mu\nu}(\nabla_\beta \phi \nabla^\beta\phi)^{(1+w_{\pm})/(2w_{\pm})}
\biggr],
\label{tmunuphi}
\end{eqnarray}
and the field equation of motion (stemming from the energy-momentum conservation)
is given by
\begin{equation}
\nabla^{\mu}\nabla_{\mu}\phi+\frac{(1-w_{\pm})}{w_{\pm}}\frac{\nabla_\mu \phi \nabla_\nu\phi
\nabla_{\alpha}\nabla_\beta \phi g^{\mu\alpha}g^{\nu\beta}}
{\nabla_\rho \phi \nabla^\rho\phi} =0 .
\label{eqscalarfield}
\end{equation}

In the homogeneous background (\ref{dg0}), the energy density and pressure  read
\begin{equation} \epsilon_{\pm (0)}\equiv T^0_{(0)0}=\pm \frac{1}{2w_{\pm}}
\left(\frac{\varphi '}{a}\right)^{(1+w_{\pm})/w_{\pm}} \label{epspm}
\end{equation}
\begin{equation}
p_{\pm (0)}\equiv -\frac{T^i_{(0)i}}{3}=w_{\pm}T^0_{(0)0}=w_{\pm}\epsilon_{\pm (0)},\label{prho}
\end{equation}
and the equation of motion (\ref{eqscalarfield}) reduces to
\begin{equation} \varphi''+(3w_{\pm} -1)\Hu\varphi'=0,
\end{equation} \noindent
where $\varphi$, $\epsilon_{\pm (0)}$ and $p_{\pm (0)}$ are the homogeneous parts of 
$\phi$, $\epsilon_{\pm}$ and $p_{\pm}$, respectively.  In (\ref{epspm}) and
the following equations, a prime represents a derivative
with respect to the conformal time $\eta$ of the metric (\ref{dg}).

From the equation for $\varphi$ one obtains $\varphi'=C_{\varphi}/a^{(3w_{\pm}-1)}$, where
$C_{\varphi}$ is a constant related to $\rho_{\pm}$ through
\begin{equation}
\rho_{\pm} = \frac{C_{\varphi}^{(1+w_{\pm})/w_{\pm}}}{2w_{\pm}}.
\end{equation}
In order to make contact with the notation of Ref.~\cite{PPN},
which is contained in the general case we treat here ($\alpha=1/2$, $w_-=1$),
we will choose to represent the negative energy perfect fluid
by this K-essence scalar field and let the positive energy fluid
with its original hydrodynamical representation, thus leading to
the action
\begin{widetext}
\begin{equation} {\cal S} = - \int \left[{1\over 16\pi G} R + \epsilon_+
+ \frac{1}{2}(\nabla_\mu \phi \nabla^\mu
\phi)^{(1+w_-)/(2w_-)}\right] \sqrt{-g}\,
\dd^4x ,\label{action0}
\end{equation}
\end{widetext}
where $R$ is the curvature scalar that takes into account 
gravity. Eq.~(\ref{action0}) reduces to that of Ref.~\cite{PPN} if $w_-=1$.

The full Einstein equations then reads
\begin{equation}
G_{\mu\nu} = 8\pi G(T^+_{\mu\nu}+T^-_{\mu\nu}),
\label{action}
\end{equation}
where $G_{\mu\nu}$ is the Einstein tensor, $T^{\pm}_{\mu\nu}$
are the energy-momentum tensors of the positive and negative energy perfect
fluid respectively, with $T^+_{\mu\nu}$ written in the hydrodynamical representation,
and $T^-_{\mu\nu}$
expressed in terms of $\phi$ given by Eq.~(\ref{tmunuphi}).

\subsection{Regular equations for cosmological perturbations.}

A set of regular equations is a necessary tool for a numerical
analysis of bounce physics. We shall generalize the treatment of
Ref. \cite{PPN} to the generalized class of bouncing models found in
Sect. II. The most general form of scalar metric perturbations on the background given
by Eq.~(\ref{sfevolution}) reads, in the longitudinal gauge,
\begin{equation} \dd s^2 = a^2(\eta) \left[ (1+2 \Phi) \dd\eta^2 -
(1-2\Phi)\delta_{ij} \dd x^i \dd x^j \right],\label{dg}\end{equation}
where $\Phi$ is the gauge invariant Bardeen
potential~\cite{bardeen,MFB}. For the matter fields we have
\begin{equation} \phi = \varphi (\eta) + \delta\phi ({\mathbf x},\eta)
\ \ \ \hbox{\rm and} \ \ \
\epsilon_{\pm} = \varepsilon_{\pm} (\eta) + \delta\epsilon_{\pm} ({\mathbf x}, \eta),
\label{dphieps}
\end{equation}
where $\epsilon_{\pm}\equiv\pm T^0_{0\pm}$.

{}From Eq.~(\ref{tmunuphi}) at first order we obtain
\begin{equation}
\delta_-\equiv \frac{\delta\epsilon_{-}}{\varepsilon_{-}}=
\frac{1+w_-}{w_-}\left(\frac{\delta\phi'}{\varphi'}-\Phi\right).
\label{rho-}
\end{equation}
One can also check from Eq.~(\ref{tmunuphi}) that
$\delta p_- \equiv \delta T^i_i /3=w_- \delta\epsilon_-$.

\begin{widetext}
Using Eq.~(\ref{eqscalarfield}) to obtain
a linear equation for $\delta\phi$, and the perturbed Einstein
equations in order to obtain an equation for the Bardeen potential
$\Phi$, after eliminating $\delta\rho_+$, yields the coupled set
of regular equations for the modes of wavenumber $k$, namely
\begin{equation}
\Phi_k'' + 3 {\cal H} \left( 1 + w_+ \right) \Phi_k' + \left[ w_+ k^2
+ 2 {\cal H}' + ({\cal H}^2-K) (1 + 3 w_+) + \frac{3}{2} {\cal H}^2
\Omega_- F \right] \Phi_k = \frac{3}{2} {\cal H}^2 \Omega_- \, F \,
\frac{\delta\phi_k'}{\varphi'},
\label{Phi_second}
\end{equation}
and
\begin{equation}
\delta\phi_k'' + {\cal H} (3w_- -1) \delta\phi_k' + k^2 w_- \delta\phi_{k} =
(1 + 3 w_-) \varphi'\Phi_k' \,,
\label{theta_second}
\end{equation}
which we wrote in full generality by including a possibly non-vanishing
curvature of the homogeneous spatial section $K$, and we have set
$F \equiv (w_+ - w_-)(1+w_-)/w_-$, and
$\Omega_- = \varepsilon_- \lP^2 a^2/\Hu ^2$.

Another way one can write Eq.~(\ref{Phi_second}) using the background equations of 
motion, which will be usefull when discussing the possible spectra, reads:
\begin{eqnarray}
&\Phi_k''& + 3 {\cal H} \left( 1 + w_+ \right) \Phi_k' + \left[ w_+ k^2
+\frac{(w_- -1)}{w_-} {\cal H}' + (1 + 3 w_+)\frac{(w_- -1)}{2w_-}{\cal H}^2 - (1 + 3 w_+)\frac{(3w_- +1)}{2w_-}K\right] 
\Phi_k =\nonumber \\ &=& \frac{3}{4w_-} \lP^2 (2w_-\rho_-)^{1/(1+w_-)} \, F \,
\frac{\delta\phi_k'}{a^2}.
\label{Phi_second2}
\end{eqnarray}
\end{widetext}

The advantage of Eqs.~(\ref{Phi_second},\ref{theta_second}) is their
general use in bounces with two hydrodynamical fluids, constituting
a set of coupled regular equations for numerical analysis of bounce physics.
We should like to use this opportunity to mention the fact that
attempts to write down two {\em uncoupled} equations for two
separate parts of the Newtonian potential, as suggested in Ref.~\cite{PPN}
are incorrect \cite{FU,batte,batte2}, as well as its use~\cite{F}.

Another set of regular equations for general two-fluid models
solely in terms of hydrodynamical variables (without using
the K-essence scalar field to describe the $w_-$ perfect fluid)
can be obtained using, together with the perturbed Einstein
equations, the perturbed energy-momentum tensor conservation
equations
\be \delta_{-\,k}' + (1+w_-)
(\theta_{-\,k}' - 3 \Phi_k') = 0 \,,
\ee
\be \theta_{-\,k}' + {\cal H}
(1 - 3 w_-) \theta_{-\,k} - k^2 \Phi_k = \frac{w_- k^2
\delta_{-\,k}}{1+w_-} \, ,
\ee
where $(\delta u_{k\,-})_i=a\partial_i\theta_{-\,k}/k^2$ [$(\delta u_{k\,-})_i$
is the perturbed $w_-$ three-velocity mode], and $\delta_- \equiv \delta
\rho_-/\rho_-$. The result reads
\begin{widetext}
\begin{equation}
\Phi_k'' + 3 {\cal H} \left( 1 + w_+ \right) \Phi_k' + \left[ w_+ k^2
+ 2 {\cal H}' + ({\cal H}^2-K) (1 + 3 w_+) + \frac{3}{2} {\cal H}^2
\Omega_- F \right] \Phi_k = \frac{3}{2} {\cal H}^2 \Omega_- \, F \,
\frac{\theta_{-\,k}' + (1-3w_-) {\cal H} \theta_{-\,k}}{k^2},
\label{Phi_second3}
\end{equation}
\begin{equation}
\theta_{-\,k}'' + {\cal H} (1-3 w_-) \theta_{-\,k}' + \left[ k^2 w_- +
(1-3w_-) {\cal H}' \right] \theta_{-\,k} = k^2 (1 + 3 w_-) \Phi_k' \,,
\label{theta_second2}
\end{equation}
where $F \equiv (w_+ - w_-)(1+w_-)/w_-$ and
$\Omega_- = \rho_- \lP^2 a^2/{\Hu}^2$, as before.
\end{widetext}

The relation between $\theta$ and $\delta\phi$ is given by,
\be
\theta_{-\,k} = k^2 \frac{\delta \phi_k}{\varphi'},
\ee
from which the system (\ref{Phi_second},\ref{theta_second}) can be recovered from 
(\ref{Phi_second3},\ref{theta_second2}) straightforwardly.

\section{Numerical Results}

The system (\ref{Phi_second},\ref{theta_second}) written in terms
of $\Phi_k$ and $Y_k = \delta \varphi \lP$
which evolve in the variable $x=\tau/\tau_0$
for the family of bounces described by
(\ref{sfevolution}) read
\begin{widetext}
\begin{eqnarray}
\label{adim22}
\displaystyle{\dd^2\Pk\over \dd x^2} + ( 1 + \alpha + \alpha \beta) {2 x\over x^2+1}
{\dd\Pk\over \dd x} +
\left[ w_+ \tilde k^2 (1+x^2)^{2 \alpha (\beta - 1)} +
\frac{4 \alpha (1- 3 \alpha + \alpha \beta)}{(1+x^2)^2
(2-3 \alpha + 2 \alpha \beta)}
\right] \Pk \nonumber \\ =
-\frac{2 \sqrt{2}
\alpha (\alpha \beta+1) (1 - \alpha +\alpha \beta)}
{2 - 3 \alpha + 2 \alpha \beta}
(1+x^2)^{\alpha (\beta - 3)}
{\dd Y_k\over \dd x}
\,,
\end{eqnarray}
for the metric perturbation, and
\bea
\label{adim2bis2}
{\dd^2Y_k\over \dd x^2} &+& (2 + \alpha \beta - 3 \alpha)
{2 x \over x^2+1}
{\dd Y_k\over \dd x} +
w_- \tilde k^2 (1+x^2)^{2 \alpha (\beta - 1)}
Y_k =
{2 \sqrt{2} \over \alpha} (1+x^2)^{3 \alpha -2 -\alpha \beta}
{\dd \Pk\over \dd x} \,.
\eea
with $\tilde k = k \tau_0 a_0^{\beta - 1}$ for the scalar field part.

In what follows below, we have solved, numerically, the set consisting of
Eqs.~(\ref{adim22}) and (\ref{adim2bis2}), setting unnormalized vacuum initial
conditions (the fact that we do not bother about the normalization here is
because we are merely interested in the transmitted spectrum), reading
\bea
\label{ic}
\Phi_{k \,, {\rm ini}} &=& {x^{- 3 \alpha (1+w_+)} \over \sqrt{\tilde k^3}}
\exp \left[ -i \frac{\tilde k \sqrt{w_+}}{1+ 2 \alpha (\beta - 1)}
x^{2 \alpha(\beta - 1) + 1}
\right],
\nonumber \\
Y_{k \,, {\rm ini}} &=& {x^{\alpha (1-3 w_-)} \over \sqrt{\tilde k}}
\exp \left[ -i \frac{\tilde k \sqrt{w_-}}{1+ 2 \alpha (\beta - 1)}
x^{2 \alpha(\beta - 1) + 1}
\right].
\eea
\end{widetext}

\begin{figure*}[t]
\includegraphics[width=8cm]{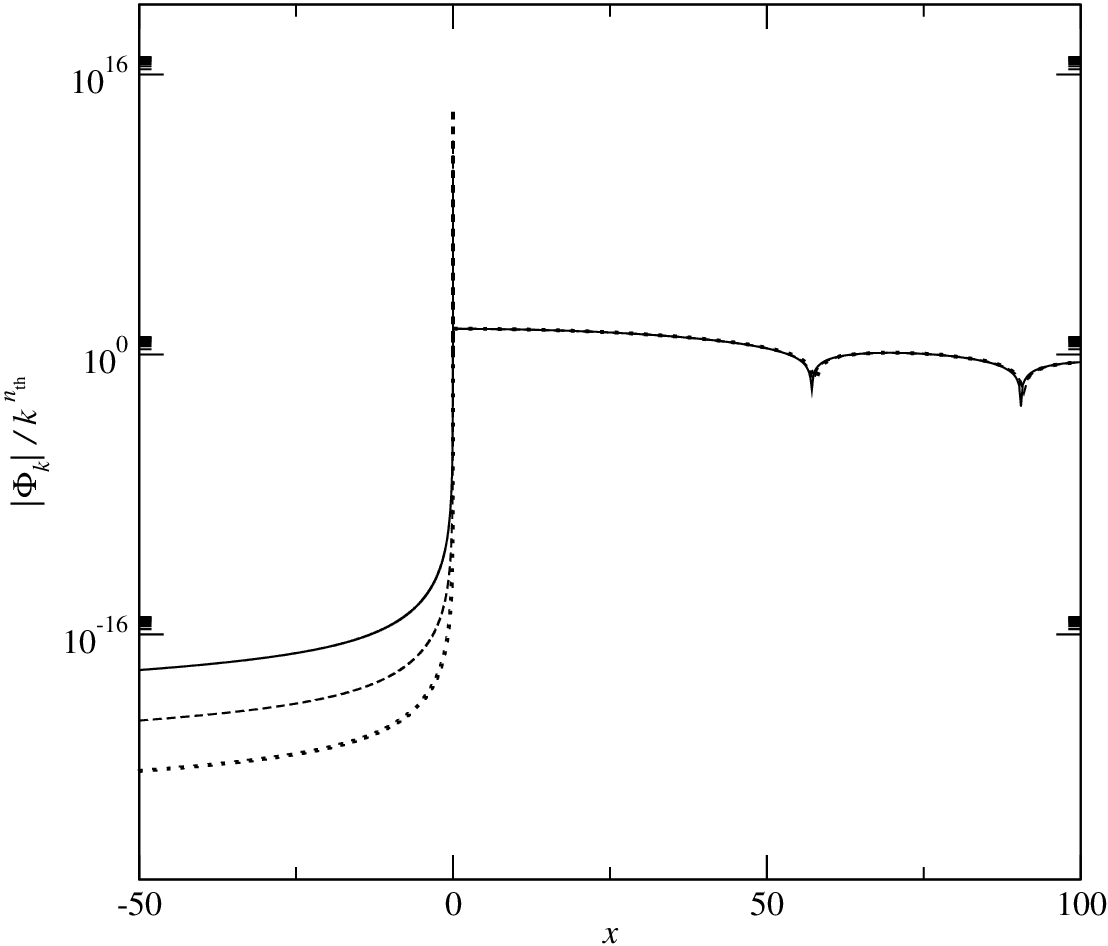}\hskip1cm
\includegraphics[width=8cm]{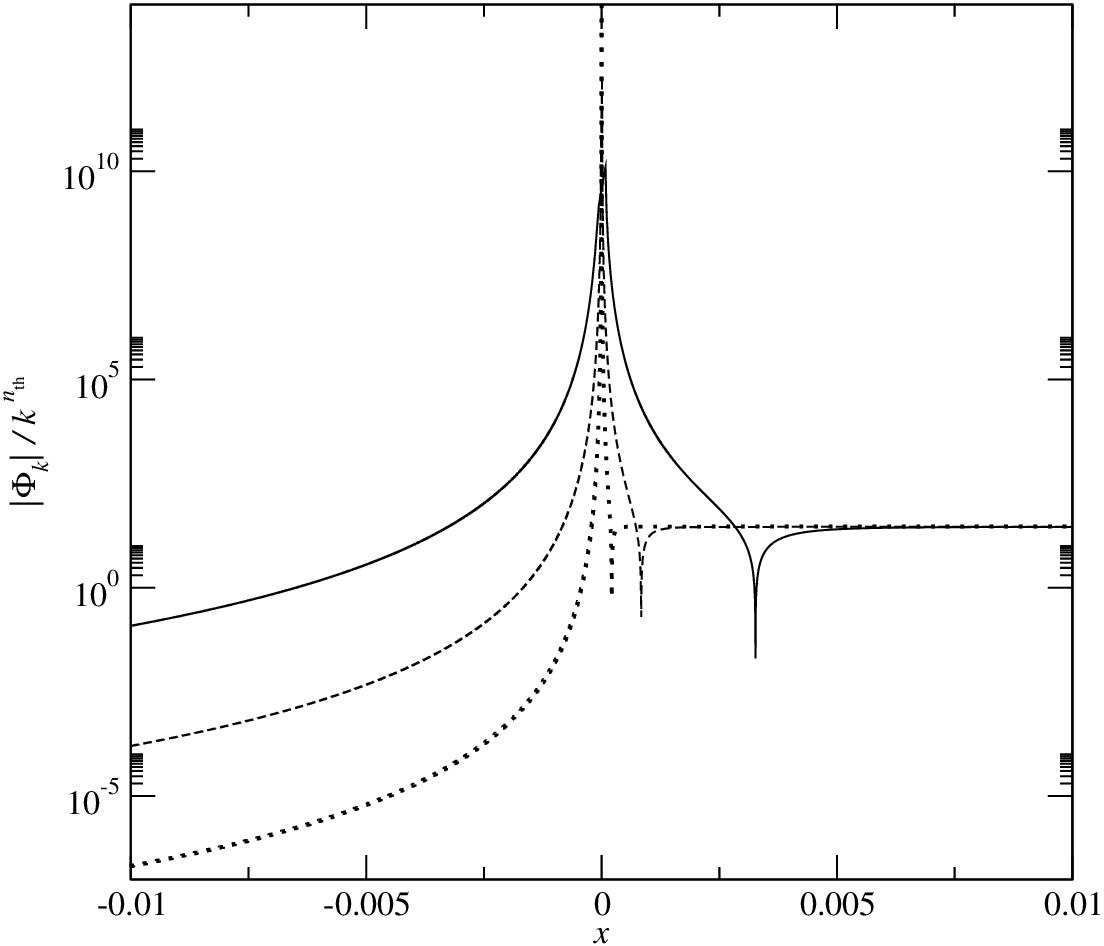}
\caption{Example of the time dependence of the Newtonian potential for
three different wavelengths ($\tilde k = 10^{-5}, 10^{-6}$ and
$10^{-7}$ respectively) as function of $x=\eta/\eta_0$. This example,
for which $w_-=1$ and $w_+=10^{-2}$ is typical of most cases for which
there is a constant mode, as found in \cite{BV1,BV2}. Note that the amplitude at
the bounce can be much larger than that of the constant mode that dominates
later. The Bardeen potential is here rescaled by the predicted spectrum, which
in this case is given by Eq.  (\ref{scalar_theory})
}
\label{fig:time}
\end{figure*}

\begin{figure}[t]
\includegraphics[width=8cm]{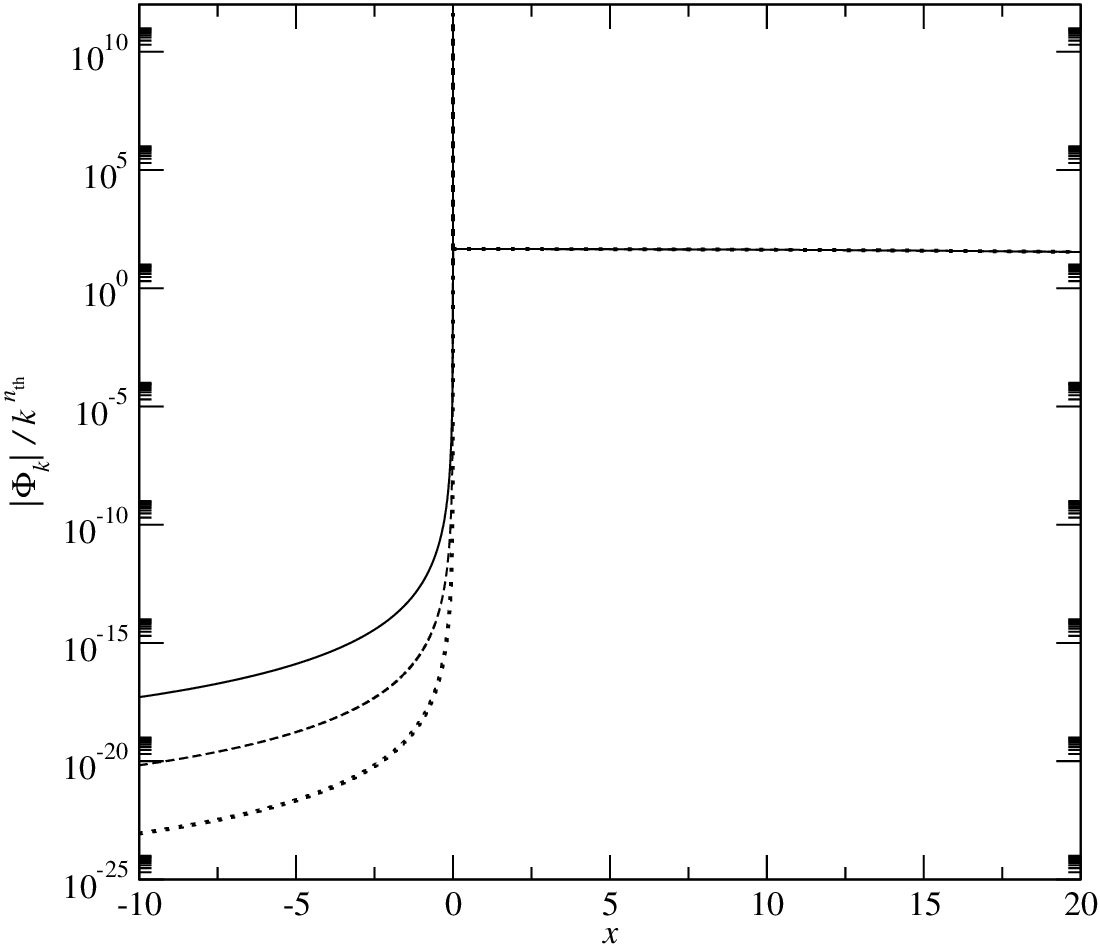}
\caption{Same as FIG.~ \ref{fig:time} with a different value  for $w_+$, namely $w_+=1/4$.
Again, the Bardeen potential is rescaled with the spectrum found by \cite{BV1,BV2}, which we thus independently confirm.  
}
\label{fig:time2}
\end{figure}

Fig.~\ref{fig:time} shows the time evolution for the spectrum
$P_\Phi /k^{n_{\rm S}-1}$, for that particular case for which the theoretical value
for the scalar spectral index $n_{\rm S}$ in the expanding stage is known
and given by Eq. (\ref{scalar_theory}), i.e. for $w_-=1$ and $w_+=10^{-2}$. The
plots for three different wavenumbers show how the spectrum for the 
(growing mode of the) Newtonian potential in the expanding stage is the 
one of curvature perturbations in the contracting stage. Fig.~\ref{fig:time2} shows
the same plot (near the bounce only) for a different value of  $w_-$, namely $w_- = 1/4$,
again rescaled with the theoretical prediction.

\section{Analytic Study Of The Stiff Matter Case}

In this section, we will analytically justify the spectra of the class of
bounces which are driven by negative energy stiff matter.
Note that for $w_-=1$ in the equation for the
Newtonian potential $\Phi_k$
(\ref{Phi_second2}), the ``mass'' term on the left hand side vanishes for flat 
spatial sections, $K=0$, and infinite wavelength, $k=0$.
In this case, one can obtain a solution around the bounce which can be
matched with the solutions far from it to obtain the spectra for large wavelengths.
This section is a generalization of what was done in Ref.~\cite{PPN}, which
concentrated on the radiation-stiff matter case only.

\subsection{The relevant phases in the perturbations evolution}

Let us first consider the asymptotic limit $\eta\rightarrow -\infty$, where
$x\propto\eta^{3(1-w_+)/(1+3w_+)}$
and $a\propto\eta^{2/(1+3w_+)}$ (the positive energy fluid dominates).
Taking into account the initial conditions (\ref{ic}), which in terms of $\eta$
read 
\bea
\label{ic2}
\Phi_{k \,, {\rm ini}} &\propto& \frac{\exp (-i k \eta)}{\eta^{3(1+w_+)/(1+3w_+)}\sqrt{k^3}},
\nonumber \\ 
X_{k \,, {\rm ini}} &\propto& \frac{\exp (-i k \eta)}{\eta^{2/(1+3w_+)}\sqrt{k}} ,
\eea
one can see that the source terms in
Eqs.~(\ref{Phi_second2} and (\ref{theta_second}) are negligible provided $w_- < 7/3$.
Defining the variables $u_k\equiv a^{3(1+w_+)/2}\Phi_k$ and
$v_k\equiv a^{(3w_- -1)/2}\delta\phi_k$, one obtains the equations:
\begin{equation}
u_k''+\left[w_+ k^2-6\frac{(1+w_+)}{(1+3w_+)^2\eta^2})\right]u_k =0,
\label{Bardeen2}\end{equation} and
\begin{equation} v_k''+\left[k^2-\frac{\left({a^{(3w_- - 1)/2}}\right)''}{a^{(3w_- - 1)/2}}\right]v_k = 0.
\label{scalar2}
\end{equation}
The solutions of these equations are 
\begin{equation} 
\label{rad27}
u_k = \sqrt{\eta}\left[ \Phi_{_{(1)}} H^{^{\rm
(1)}}_{\nu} (\sqrt{w_+}k\eta) + \Phi_{_{(2)}} H^{^{\rm (2)}}_{\nu}
(\sqrt{w_+}k\eta) \right], 
\end{equation}
from which one derives $\Phi_k$, and
\begin{eqnarray} 
\delta\phi_k &=& a^{(1-3w_-)/2} \sqrt{\eta}[ X_{_{(1)}} H^{^{\rm (1)}}_{\mu} (\sqrt{w_-}k\eta) 
\nonumber \\ &+& 
X_{_{(2)}} H^{^{\rm (2)}}_{\mu} (\sqrt{w_-}k \eta)].
\label{radiation}
\end{eqnarray}
where $\nu=\frac{5+3w_+}{2(1+3w_+)}$ and $\mu=\frac{3+3w_+ - 6w_-}{2(1+3w_+)}$.
The coefficients $\Phi_{_{(i)}}$ and $X_{_{(i)}}$ are time-independent and only depend
on $k$.

Restricting now to the case $w_-=1$, and taking into account the initial conditions
(\ref{ic2}), one obtains that $\Phi_{_{(1)}}=X_{_{(1)}}=0$, $\Phi_{_{(2)}}=1/k$,
and $X_{_{(2)}}=1$. In the region where $k\eta \ll 1$ (we are considering long wavelengths)
but still far from the bounce,
where the source terms can still be neglected, one has
\begin{equation}
\Phi_{<\;k} = A_1+\frac{A_2}{\eta ^{(5+3w_+)/(1+3w_+)}}+ O(k^2\eta^2),
\label{phi2}
\end{equation}
and
\begin{equation}
\delta\phi_{<\;k}=B_1 + \frac{B_2}{\eta ^{3(1 - w_+)/(1+3w_+)}}+ O(k^2\eta^2),
\label{sca3}
\end{equation}
where $A_1=k^{3(1-w_+)/[2(1+3w_+)]}$, $A_2=k^{(-7-9w_+)/[2(1+3w_+)]}$, $B_1=k^{3(1-w_+)/[2(1+3w_+)]}$
and $B_2=k^{3(w_+-1)/[2(1+3w_+)]}$.

Now we have to propagate this solution through the bounce and match it with
the solution in the expanding phase. During the bounce, the source terms 
in Eqs.~(\ref{Phi_second2}) and (\ref{theta_second})
cannot be neglected but now it is the terms proportional to $k^2$ 
that are negligible.
For $w_-=1$, Eq.~(\ref{Phi_second2}) simplifies and, upon returning 
to the variables $Y\equiv \lP\delta\phi$ and $x=\tau/\tau_0$,
one obtains 
\begin{equation}
\label{adim2}
{\dd^2\Pk\over \dd x^2} + {8\alpha x\over x^2+1}
{\dd\Pk\over \dd x} = -\frac{\sqrt{2}}{(1+x^2)}
{\dd Y_k\over \dd x}
\,,
\end{equation}
\be
\label{adim2bis}
{\dd^2Y_k\over \dd x^2} + {2 x \over x^2+1}
{\dd Y_k\over \dd x} =
{8 \sqrt{2} \alpha \over (1+x^2)}
{\dd \Pk\over \dd x} \,,
\ee
where, as $w_-=1$, $\alpha=1/[3(1-w_+)]$.
The solutions of these equations read
\begin{equation} 
\Phi^{^{\rm Bounce}} = \tilde{A} + \tilde{B} f_1(x)+\tilde{C} f_2(x),
\label{Phi0}
\end{equation}
\noindent and
\begin{equation} 
Y{^{\rm Bounce}} = \tilde{D} + \tilde{B}f_3(x)+ \tilde{C} f_4(x), 
\label{X0}
\end{equation}
\noindent with $\tilde{A}$, $\tilde{B}$, $\tilde{C}$ and $\tilde{D}$
arbitrary constants. The bounce functions $f_i(x)$ are found to be
\begin{widetext}
\begin{equation} 
f_1 (x) \equiv {x\over (1+x^2)^{4\alpha}}\; , \;\;\; f_3 (x) \equiv -
\frac{\sqrt{2}} {(1+x^2)^{4\alpha}}
\end{equation}
\begin{equation}  
f_2(x)\equiv -8\alpha\int^x {\rm d}{\tilde {x}}\left\{\frac{{\tilde {x}}}{(1+{\tilde {x}}^2)^{1+4\alpha}}\left[({\tilde {x}}^2+1) \, 
F\left(\frac{1}{2},-4\alpha +1,\frac{3}{2},-{\tilde {x}}^2\right)
+{\tilde {x}}F\left(-\frac{1}{2},-4\alpha,\frac{1}{2},-{\tilde {x}}^2\right)\right]\right\}
\end{equation}
\begin{equation} 
f_4(x)\equiv \int^x {\rm d}{\tilde {x}}
\left[\frac{1}{(1+{\tilde {x}}^2)^{1+4\alpha}} 
F\left(-\frac{1}{2},-4\alpha,\frac{1}{2},-{\tilde {x}}^2\right)\right],
\end{equation}
where $F$ denotes the hypergeometric function.
For $x\gg 1$, the solutions can be written as
\begin{equation} 
\Phi^{^{\rm Bounce}} \approx \tilde{A} + \frac{\tilde{B}}{x^{8\alpha-1}} +\frac{\tilde{C}}{x^2}
= \tilde{A} + \frac{\tilde{B}}{\eta^{(5+3w_+)/(1+3w_+)}} +\frac{\tilde{C}}{\eta^{6(1-w_+)/(1+3w_+)}},
\label{Phi01}
\end{equation}
\noindent and
\begin{equation} 
Y{^{\rm Bounce}} \approx \tilde{D} - \frac{\sqrt{2}\tilde{B}}{x^{8\alpha}}+ \frac{\tilde{C}}{x}
= \tilde{D} + \frac{\tilde{B}}{\eta^{8/(1+3w_+)}} +\frac{\tilde{C}}{\eta^{3(1-w_+)/(1+3w_+)}}. 
\label{X01}
\end{equation}
These solutions coincide with those obtained in Ref.~\cite{PPN} for $w_+=1/3$.
\end{widetext}

If we now compare Eqs.~(\ref{Phi01},\ref{X01}) with Eqs.~(\ref{phi2},\ref{sca3}), one can obtain that
$\tilde{A}=k^{3(1-w_+)/[2(1+3w_+)]}$, $\tilde{B}=k^{(-7-9w_+)/[2(1+3w_+)]}$, $\tilde{D}=k^{3(1-w_+)/[2(1+3w_+)]}$
and $\tilde{C}=k^{3(w_+-1)/[2(1+3w_+)]}$. One can also see this by noting that the third term in Eq.~(\ref{Phi01}) 
is the first contribution of the source term to $\Phi$, which of course must have the $k$-dependence of $B_2$ in
(\ref{sca3}), while the second term of Eq.~(\ref{X01}) is the first contribution of the source term to $\delta\phi$, 
which must have the $k$-dependence of $A_2$ in (\ref{phi2}).

We now have to match Eqs.~(\ref{Phi01},\ref{X01}) with the solutions in
the expanding phase which are far from the bounce and where the source terms are negligible, i.e.:
\begin{equation}
\Phi_{>\;k} = {\bar{A}}_1+\frac{{\bar{A}}_2}{\eta ^{(5+3w_+)/(1+3w_+)}}+ O(k^2\eta^2),
\label{phi8}
\end{equation}
and
\begin{equation}
\delta\phi_{>\;k}={\bar{B}}_1 + \frac{{\bar{B}}_2}{\eta ^{3(w_+ - 1)/(1+3w_+)}}+ O(k^2\eta^2),
\label{sca8}
\end{equation}
yielding, for the constant part of $\Phi$, which determines the spectrum,
\begin{eqnarray}
{\bar{A}}_1&\propto & k^{3(1-w_+)/[2(1+3w_+)]}+k^{3(w_+-1)/[2(1+3w_+)]}
\nonumber \\&\approx &k^{3(w_+-1)/[2(1+3w_+)]},
\end{eqnarray}
because $w_+ < 1$ and we assume $k\ll 1$. Note that $\Phi$ gets the spectrum of $\delta\phi$.

If we now calculate the power spectrum
\begin{equation}
{\cal P}_k\equiv \frac{k^3}{2 \pi^2}
|\Pk\,\,|^2 \equiv \AS k^{\nS-1},
\label{spectrum}
\end{equation}
we obtain
\begin{equation}
\nS -1=\frac{12w_+}{1+3 w_+},
\label{scalar_theory}
\end{equation}
as obtained in the numerical calculations and in Refs.~\cite{BV1,BV2}.
This result was also obtained by considerations on the matching 
conditions in \cite{DM,BF,FB}, which predict the spectral index in the 
expanding stage as the one of curvature perturbations in the contracting 
stage.
Note incidentally that it coincides with the spectrum obtained in Ref.~\cite{PPP},
where the bounce is not caused by a negative energy stiff matter but by quantum effects:
the background and the spectrum have the same behavior.

\section{Gravitational Waves}

The equation for the Fourier transforms of the amplitude of the 
two polarization degrees of 
gravitational waves in cosmology is
\be
\frac{\dd^2 {h_k}}{\dd t^2} +
3 H \frac{\dd {h_k}}{\dd t}
+ \frac{k^2}{a^2} h_k=0 \,.
\label{eq_motion}
\ee
Once we introduce $v\equiv a^{(3-\beta)/2} h$ and use the same coordinate 
variable $\tau$ as introduced in Eq. (2), the above equation becomes:
\begin{widetext}
\be
\ddot v_k +
\left[ k^2 a^{2(\beta-1)} + \frac{(\beta-3)}{2} \frac{\ddot 
a}{a} -\frac{(\beta-1)(\beta-3)}{4} \left(\frac{\dot a}{a}\right)^2 
\right] v_k=0,
\label{eq:v}
\ee
where $\dot f \equiv \dd f/\dd \tau$.
The very existence of an asymptotic vacuum demands the condition
$2\alpha(1-\beta)<1$, or, in other words, if $w_+>-1/3$,
which we shall therefore assume. From here on, for convenience, we also define
$\delta\equiv 1+2\alpha (\beta-1) >0$. We also restrict to $w_+ < 1$ 
in order to have 
the constant mode as the growing mode in the expanding stage.

The equation which we numerically evolve is:
\be
\frac{\dd^2 v_k}{\dd x^2} + \left[ \tilde k^2 
\left( 1 + x^2\right)^{2\alpha (\beta-1)} 
+  \frac{\alpha(\beta-3)}{\left( 1+x^2 \right)^2} \left\{ 1-\left[1+\alpha
(\beta-3)\right]x^2 \right\} \right] v_k=0
\label{gw_x}
\ee
with, as usual, $x=\tau/\tau_0$, $\tilde k = k a_0^{\beta-1} \tau_0$.

Finally, the initial conditions corresponding to the adiabatic vacuum are
taken to be
\be
\mu_\mathrm{k \,, ini} = \displaystyle \frac{\sqrt{3}\lP}{\sqrt{k}} 
\exp (-ik\eta) \ \ \Longrightarrow \ \ \ 
v_\mathrm{k \,, ini} =\sqrt{3\tau_0}\lP \frac{\displaystyle 
x^{\alpha(1-\beta)}}{\sqrt{\tilde k}}
\exp (-i k\eta),
\label{mu:ini}
\ee
where $\mu_k = a\, h_k$, and we get rid of the prefactor since we are
mostly interested in the spectral index anyway (just like for the scalar case,
the normalisation here is essentially irrelevant). Recall also that
$k\eta = \tilde k x^\delta/\delta.$

\subsection{Analytic Approximations}

We are first interested in determining the matching point between the short and long wavelength approximations.
The potential in terms of the conformal time is
\be
\frac{a''}{a} = a^{2(1-\beta)} \left[ \frac{\ddot a}{a} + (1-\beta)\left(
\frac{\dot a}{a}\right)^2\right] \equiv V(\eta),
\ee
Expliciting this in the $\tau$ variable, this is
\be
V=2\alpha a_0^{2(1-\beta)} f(\tau),
\ee
where
\be
f(\tau)\equiv -\left\{ \left[ 1+2\alpha \left(\beta-2\right)\right] \tau^2-\tau_0^2 \right\}\tau_0^{4\alpha(\beta-1)} 
\left(\tau^2+\tau_0^2\right)^{-2[1+\alpha(\beta-1)]}
\ee
so the matching point at which $k^2\sim |a''/a|$, i.e
\be
x_{_\mathrm{M}} = \left\{ \frac{\tilde k}{\sqrt{2\alpha \left[1+2\alpha(\beta-2) \right]} }\right\}^{-1/\delta}\gg 1,\label{xM}
\ee
where the last inequality stems from the requirement that there is an asymptotic vacuum, i.e.
$\delta > 0$.

The zeros of the first derivative of $V$ are determined by the equation
\be
\tau\{[2\alpha(4\alpha -3 -6\alpha\beta + 2\beta + 2\alpha\beta^2) +1] \tau^2+[2\alpha(3-2\beta)-3]\tau_0^2\}=0.
\ee 
\end{widetext}
We will here treat the simplest case where the potential $V$ has only one extremal point, at $\tau =0$,
hence imposing that the coefficients of $\tau^2$ and $\tau^2_0$ have the same sign.

\begin{figure}[t]
\includegraphics[width=8cm]{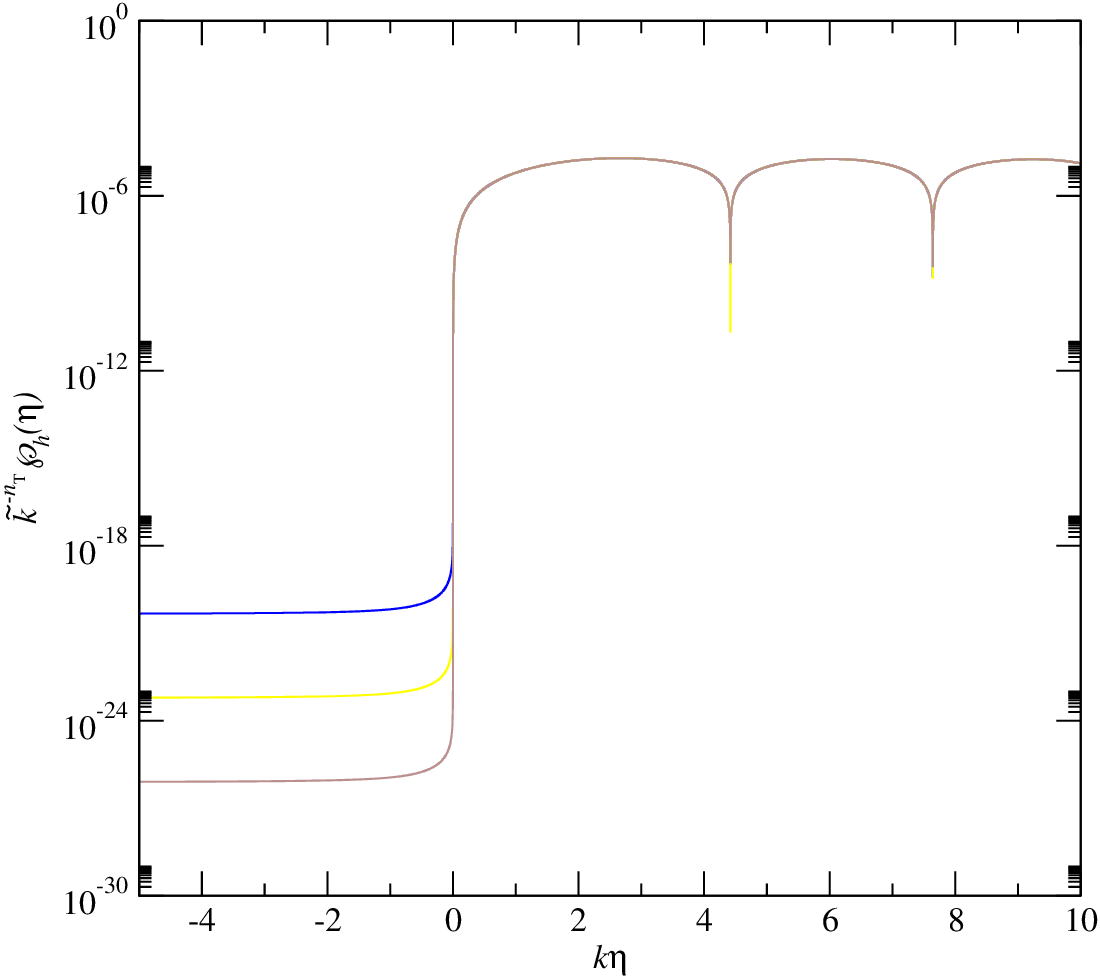}
\caption{Example of the time dependence of the tensor perturbation for
three different wavelengths ($\tilde k = 10^{-5}, 10^{-6}$ and
$10^{-7}$ respectively) as function of $k\eta$. This example,
for which $w_-=1/3$ and $w_+=10^{-2}$ is typical, again, and in
complete agreement with our analytical prediction. All other cases
lead to identical figures, except for the actual numbers. The cases
for which the potential does not satisfy the condition that the
potential has only one extremum at $\tau=0$ cannot be
compared with theoretical expectation, in the lack of it, and so are
not shown.}
\label{fig:Grav}
\end{figure}

Asymptotically far from the bounce, Eq. (\ref{gw_x}) becomes:
\be
\frac{\dd^2 v_k}{\dd x^2} + 
\left[ \tilde k^2 x^{4\alpha(\beta-1)} - 
\frac{\gamma(1+\gamma)}{x^2}\right] v_k=0,
\label{eq:v_asym}
\ee
where $\gamma\equiv \alpha(\beta-3)$; the above equation admits a solution in terms of 
the Hankel function, in accordance with the vacuum initial conditions (\ref{mu:ini}):
\be
v_\mathrm{k \,,1} = A \sqrt{x} H^{(2)}_\nu \left(\frac{\tilde k 
x^\delta}{\delta}
\right), \label{v1}
\ee
supposed to be valid up to $x_{_\mathrm{M}}$ of Eq.~(\ref{xM}), 
with $$ A^2 = \frac{3\pi\tau_0}{2\delta}
\lP^2\exp^{-i\pi(\nu+\frac{1}{2})} \,,$$ and
$$\nu\equiv \frac{\gamma+\frac{1}{2}}{\delta} .$$
Note incidentally at this point that the matching time (\ref{xM}) gives an argument for the
Hankel function which does not depend on $\tilde k$. We shall henceforth call $H_\nu^{(2)} (\tilde k
x_{_\mathrm{M}}^\delta/\delta ) = H_\nu^{(2)} \left( 1/\delta\right) \equiv h^{\scriptscriptstyle{M}}_\nu$.

On the other hand, for long wavelengths close to the bounce, Eq.~({\ref{gw_x}) simplifies to
\be
\frac{\dd^2 v_k}{\dd x^2} 
+\frac{\gamma\left[1-\left(1+\gamma\right)x^2\right]}{\left(1+x^2\right)^2} 
v_k=0.
\ee
In this limit, setting $v_k=\sqrt{1+x^2}u$ and $z=ix$,
one gets the Legendre equation
\be
(1-z^2) \frac{\dd^2 u}{\dd z^2} -2z\frac{\dd u}{\dd z} + \left[ \gamma (\gamma+1)-\frac{(1+\gamma)^2}{1
-z^2} \right] u=0,
\ee
which in this case has, as the two independent solutions, a power law and a hypergeometric function.
Summarizing, we obtain, in this second regime, the general solution
\bea
v_\mathrm{k \,, 2} &=& \left( 1+x^2\right)^{-\gamma/2} 
\left[ B + C x\,  {}_2F_1\left(\frac12,-\gamma,\frac32,-x^2\right) \right]
\cr &\sim& x^{-\gamma} \left[ B + C \frac{\sqrt{\pi}\Gamma
\left(-\frac12-\gamma\right)}{2\Gamma(-\gamma)}\right] +
\frac{C x^{\gamma+1}}{1+2\gamma} + \cdots \cr & & \label{v2}
\eea

It is now a simple matter to match the solutions (\ref{v1}) and (\ref{v2}) as well as their
derivatives to get
\be
A h^{\scriptscriptstyle{M}}_\nu \tilde k^{-1/(2\delta)} = (B+C\Upsilon)  \tilde k^{\gamma/\delta} +
\frac{C}{1+2\gamma} \tilde k^{(-1-\gamma)/\delta},
\ee
and
\bea
& &\frac{A}{2} \tilde k^{1/(2\delta)}\left(h^{\scriptscriptstyle{M}}_\nu + h^{\scriptscriptstyle{M}}_{\nu-1}
-h^{\scriptscriptstyle{M}}_{\nu+1} \right) 
\nonumber \\
&=& (-C \Upsilon -B )\gamma
\tilde k^{(\gamma+1)/\delta} + \frac{C}{1+2\gamma} \tilde k^{-\gamma/\delta},
\eea
where we have set
$$
\Upsilon  \equiv \frac{\sqrt{\pi} \Gamma \left( -\frac12-\gamma\right)}{2\Gamma\left( -\gamma\right)}
$$
for notational simplicity. The solution of this system provides $B$ and $C$ a functions of the
reduced wavenumber $\tilde k$, and we shall retain in what follows the leading order terms, which
is, as we are considering $w_+ <1$, $\tilde k^{(1+2\gamma)/(2\delta)}=\tilde k^{3(w_+-1)/[2(3w_++1)]}$,
yielding for $h\approx x^{\gamma}\mu$,
\be
h\approx \tilde k^{3(w_+-1)/[2(3w_++1)]}({\rm const.} + x^{2\gamma+1}).
\ee

The actual gravitational wave spectrum is
\be
\mathcal{P}_h \equiv \frac{2 k^3}{\pi^2} \left| h \right|^2,
\ee
so we end up with
\be
\mathcal{P}_h\propto \tilde{k}^{n_\mathrm{T}},
\ee
being
\be
n_{_\mathrm{T}} = \frac{12 w_+}{1 + 3 w_+}  = \frac{2\alpha}{1+2\alpha(\beta-1)}.
\label{nT}
\ee
It is worth pointing out at this stage that Eq.~(\ref{nT}) gives the same result
as in the scalar case [Eq.~(\ref{scalar_theory})] for the specific case that we could study
analytically. The reason for such similar results stems from the fact that the dominant terms
which match through the bounces under investigation are the growing modes of curvature
perturbation and gravitational waves, both satisfying the same differential equation in the
single fluid regime. The above result  (\ref{nT}) was already obtained in previous investigations
\cite{F,PPP}, although for two different subsets of the family of bounces studied here.

\section{Conclusions}

In all the early universe models which aim at solving the horizon problem with a contraction
instead of a superluminal expansion, a deep understanding of the physics at the bounce is
crucial (and presently lacking in its full generality).
What we have shown here is a step towards the understanding of cosmological perturbations
through a bounce triggered by a second perfect fluid (with negative energy density), in the
framework of flat spatial section and general relativity. 

We have analysed in greater details, both numerically and analytically, this class of two perfect
fluid bounces with flat spatial sections, using a completly regular system of equations,
concluding indeed that the constant mode
of the scalar gravitational potential after the bounce does not acquire a piece
of the growing mode before the bounce. Therefore, our conclusions
agree with \cite{BV1,BV2}, and are in contrast with our previous 
results for scalar perturbations in \cite{PPN,F}.
Another important result is the unsensitivity of the 
scalar spectral index from the peculiarities of the bouncing component in 
the class of models studied in this paper.
One interesting result is that when the
negative energy fluid has stiff matter equation of state, the background model
and the perturbations have the same behaviour as the quantum bouncing 
cosmological
models analyzed in Ref.~\cite{PPP}.
Our results are interesting for the predictions of cosmological 
alternative models. Whereas by a very slow 
contraction - as in Ekpyrotic/cyclic model - it seems really difficult to 
generate a nearly scale-invariant spectrum of curvature perturbations 
without the need of isocurvature perturbations or extra-dimensions, a homogeneous
dust contraction \cite{FB,PPP} seems in agreement with observations and 
even free from details due to the bouncing component which 
were left open from previous investigations which focuses on 
$w_-=1$ \cite{PPP,AW}. Note however, as far as complete model building is
concerned, the assumption of homogeneity may not hold close to the bounce and
should thus be verified. This point is
however out of the scope of the present article whose aim was to concentrate on
the propagation of scalar and tensor perturbations through a regular, although
phenomenological, bounce. 

We have also performed the analytical and numerical calculations for 
gravitational waves. 
In this case, the constant mode
of the long wavelength tensor perturbations after the bounce do acquire 
a piece of the growing mode before the bounce. Also in this 
case, the slope of the final spectrum does not depend
on the negative energy perfect fluid equation of state. This paves the
way to a generic behavior for tensor perturbations, as such a phenomenological
model thus does not suffer from the drawback (still present for the scalar modes)
of relying heavily upon the details of the bounce physics.
Both these results agree with the previous investigation \cite{F} for 
a restricted class of bounces.
This can be understood by noticing that
the crucial time in the evolution of the perturbations is when
the perturbation wavelength becomes comparable with the curvature
scale of the background, when, for large wavelengths, the universe
is still far from the bounce and hence the effects of the negative energy
fluid are negligible. This result was already anticipated in Ref.~\cite{PPP}.

\section{Acknowledgements}

P. P. and N. P.-N. wish to thank CNPq of Brazil for financial support.
We also would like to thank CAPES (Brazil) and COFECUB (France) for 
partial financial support.
F. F. is partially supported by INFN BO11 and PD51. We would
also like to thank both the Institut d'Astrophysique de Paris and the
Centro Brasileiro de Pesquisas F\'{\i}sicas, where part of this work was
done, for warm hospitality and partial support (FF). We very gratefully acknowledge
various enlightening conversations with Robert Brandenberger,
J\'er\^ome Martin and David Wands. Special thanks are due to Valerio Bozza
and Gabriele Veneziano for their careful reading of the manuscript and
constructive remarks. We also would like to thank CAPES (Brazil) and 
COFECUB (France) for partial financial support.

\section{Appendix: regular equations for general two-fluid
models in terms of hydrodynamical variables.}

Another possible set of regular equations uses the density contrast
$\delta_- \equiv \delta \rho_-/\rho_-$ of the fluid driving the bounce
instead of its velocity potential. The equations are:
\begin{widetext}
\begin{equation}
\Phi_k''+ 3 {\cal H} \left( 1 + w_+ \right) \Phi_k' + \left( w_+ k^2 +
2 {\cal H}' + {\cal H}^2 + 3 w_+{\cal H}^2 \right) \Phi_k= \frac{3}{2}
{\cal H}^2 \delta_- \Omega_- (w_+ - w_-)
\label{delta-}
\end{equation}
and
\begin{eqnarray}
&\delta_- ''& + (1-3w_-)\mathcal{H} \delta_-' + \left[w_- k^2 -
\frac{9}{2} \mathcal{H}^2 (w_+ - w_-)(1+w_-) \Omega_- \right] \delta_-
=\nonumber \\ &-& (1+w_-) \left\{ \left[ k^2(1+3w_+) +3[2\mathcal{H}' + 
(1+3
w_+)\mathcal{H}^2 \right] \Phi_k + 3 [2+3(w_+ + w_-)] \mathcal{H}
\Phi_k' \right\}
\label{second_set}
\end{eqnarray}
We note that with this new set the order of the system of linear
differential equations is increased with respect to the systems
(\ref{Phi_second},\ref{theta_second}) or
(\ref{Phi_second2},\ref{theta_second2}): Eq. (\ref{second_set}) is indeed
equivalent to a third order differential equation for $\delta
\phi_k$ [see Eq.~(\ref{rho-})].
As a result, solving this last set of equations may lead to
spurious solutions and it is therefore better to stick with
Eqs. (\ref{Phi_second},\ref{theta_second}).
\end{widetext}

\end{document}